# Spin Splitting Tunable Optical Bandgap in GdN Thin Films for Spin Filtering

G. L. S. Vilela[1,2], G. M. Stephen[3,4], X. Gratens[5], G. D. Galgano[5,6], Yasen Hou[1], Y. Takamura[7], D. Heiman[1,3], A. Henriques[5], G. Berera[8] and J. S. Moodera[1,9]

[1] Plasma Science and Fusion Center, and Francis Bitter Magnet Laboratory, Massachusetts Institute of Technology, Cambridge, MA 02139, USA
[2] Física de Materiais, Escola Politécnica de Pernambuco, UPE, Recife, PE 50720-001, Brazil
[3] Department of Physics, Northeastern University, Boston, MA 02115, USA
[4] Laboratory for Physical Sciences, College Park, MD, 20740, USA
[5] Instituto de Física, Universidade de São Paulo, São Paulo, SP 05508-090, Brazil
[6] Departamento de Física, Universidade Federal de Ouro Preto, Ouro Preto, MG, 35402-136, Brazil
[7] School of Engineering, Tokyo Institute of Technology, 2-12-1 Ookayama, Meguro-ku, Tokyo, 152-8550, Japan
[8] Department of Materials Science and Engineering, Massachusetts Institute of Technology, Cambridge, MA 02139, USA
[9] Department of Physics, Massachusetts Institute of Technology, Cambridge, MA 02139, USA

E-mail: gilvania.vilela@upe.br; Moodera@mit.edu; Heiman@neu.edu

**Abstract**

Rare-earth nitrides, such as gadolinium nitride (GdN), have great potential for spintronic devices due to their unique magnetic and electronic properties. GdN has a large magnetic moment, low coercitivity and strong spin polarization suitable for spin transistors, magnetic memories and spin-based quantum computing devices. Its large spin splitting of the optical bandgap functions as a spin-filter that offers the means for spin-polarized current injection into metals, superconductors, topological insulators, 2D layers and other novel materials. As spintronics devices require thin films, a successful implementation of GdN demands a detailed investigation of the optical and magnetic properties in very thin films. With this objective, we investigate the dependence of the direct and indirect optical bandgaps ($E_g$) of half-metallic GdN, using the trilayer structure AlN(10 nm)/GdN($t$)/AlN(10 nm) for GdN film thickness $t$ in the ranging from 6 nm to 350 nm, in both paramagnetic (PM) and ferromagnetic (FM) phases. Our results show a bandgap of 1.6 eV in the PM state, while in the FM state the bandgap splits for the majority (0.8 eV) and minority (1.2 eV) spin states. As the GdN film becomes thinner the spin split magnitude increases by 60%, going from 0.290 eV to 0.460 eV. Our results point to methods for engineering GdN films for spintronic devices.

## 1. Introduction

Spintronic devices based on rare-earth nitrides (RENs) are of great interest since they present both semiconducting and ferromagnetic properties, making these materials suitable for exploiting the spin of carriers in fundamental and applied research [1-2]. Among many RENs, gadolinium nitride (GdN) is especially promising for implementations that aim to explore the electron's spin to store, process and transmit information. The advantageous properties of GdN are its high degree of spin polarization, half-metallic behavior that is useful for spin-filtering, sizeable ferromagnetic transition temperature of ~ 70 K, and large magnetic moment of $7\mu_B$/Gd$^{3+}$ due to fully occupied 4f electronic states [3-9]. The use of this material in selective transmission of electrons based on their spin orientation finds applications in magnetic tunnel junction, spin valves, spin transistors, and spin-based quantum computing, where the electron's spin state encodes and processes quantum information. In this context, the spin-filtering ability promotes an efficient way to inject and manipulate the spin current [10-15].

Due to its high oxophilicity, GdN easily reacts in air forming $Gd_2O_3$, which can cause significant changes in its electronic and magnetic properties [16-17]. The growth of stoichiometric GdN thin films, with low levels of oxygen, is very challenging and requires a rigorous control of the synthesis. This challenge has led to some disagreement in the literature regarding the electronic properties of GdN films, but most results point to its semiconducting behavior with an average bandgap of 1.2 to 1.3 eV at room temperature [5, 18-19]. In addition to controlling the stoichiometry, in order to combine GdN with novel materials that are aimed at investigating interface and spin phenomena, it is crucial to obtain atomically flat surfaces with a high degree of crystallinity, large magneto-optical effects, thermal stability and tunable bandgap energy. These properties depend on the film's strain, thickness, nitrogen vacancies, amount of oxygen impurities, parameters that depend on growth and post-growth processes, as well as protective layers and substrates choices.

In this work, we investigate the optical and magnetic properties of a set of GdN thin films with thickness ranging from $t$ = 6 nm to 350 nm. The AlN(10 nm)/GdN($t$)/AlN(10 nm) structures were reactively sputtered on sapphire (0001) substrates in a UHV sputtering system, grown at 700 °C in a pressure of 2.8 mTorr (mixture of 40% $N_2$ and 60% Ar). Measurements using a superconducting quantum interference device (SQUID) magnetometer confirm that the GdN films are magnetic with magnetization saturation of 6 $\mu_B$/$Gd^{3+}$ and a Curie temperature of ~ 30 K, in agreement with previous reports, and X-ray diffraction data confirms a good crystallinity [5,19-20]. Bandgap energies as a function of the films' thicknesses are determined from detailed analysis of the absorption spectra via extrapolated Tauc plots. The bandgap energy in both paramagnetic (PM) and ferromagnetic (FM) phases reveal a direct bandgap of ~ 1.6 eV for the upper spin-split level at 300 K in the PM state. Note that this value is significantly different from that obtained from analysis of the direct absorption coefficient and is more accurate because the bandgap is determined from extrapolation at higher energies. In the low-temperature FM state a well-separated spin-split band energy of $\Delta E = E_\uparrow - E_\downarrow \approx 0.4$ eV is found, which confirms a giant optical spin splitting in the GdN films, where $E_\uparrow$ and $E_\downarrow$ correspond to the bandgaps of the majority and minority states. We also observe the splitting magnitude depends strongly on the GdN film thickness, going from 0.290 eV for a 350 nm thick film to 0.460 eV for a 6 nm thick film, an increase of about 60%. This is an important result for incorporating GdN thin films into spintronics devices.

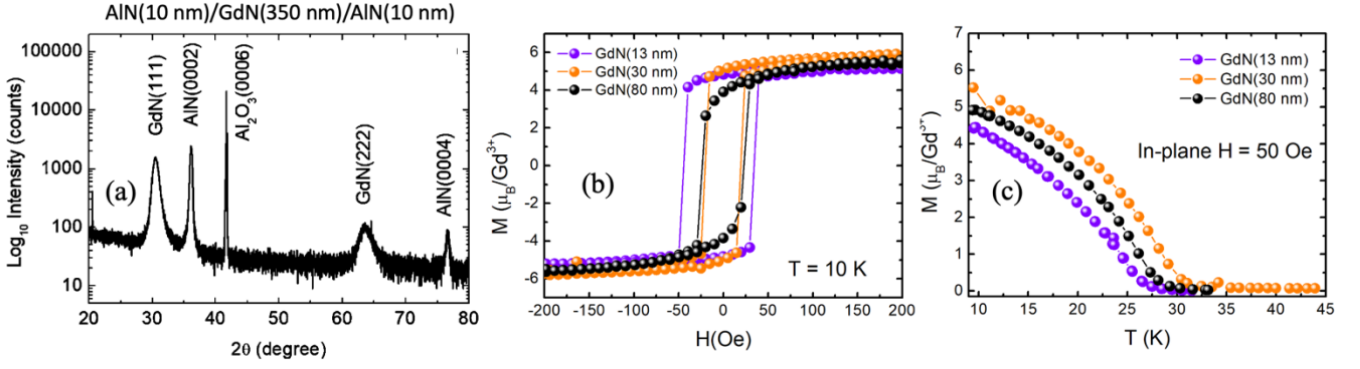

**Figure 1.** (a) High resolution X-ray diffraction pattern for the RF-sputtered AlN(10 nm)/GdN(350 nm)/AlN(10 nm) structure on a (0001)-oriented sapphire ($Al_2O_3$) substrate. (b) Magnetization ($M$) versus in-plane applied magnetic field ($H$) at a temperature of $T$=10 K for AlN(10 nm)/GdN($t$)/AlN(10 nm) structures, with $t$ = 13, 30 and 80 nm. (c) Magnetization versus temperature with an in-plane field of 50 Oe. The threes samples exhibit a Curie temperature near 30 K.

## 2. Sample Preparation, Crystal Structure and Magnetic Properties

GdN thin films with thickness ranging from 6.3 nm to 350 nm were deposited on (0001)-oriented sapphire substrates, with buffer and capping layers of 10 nm-thick aluminum nitride (AlN) films to avoid oxidation. The growth process of AlN and GdN took place in an ultra-high vacuum (UHV) sputtering chamber with a base pressure below $5 \times 10^{-8}$ Torr. To improve the substrate's surface crystallinity and film quality, prior to deposition the sapphire substrates were annealed in a quartz-tube oven at **1100** °C for 8 hours in an $O_2$ atmosphere.

After the annealing treatment, the substrates were cleaned in an ultrasonic tank first in acetone, and then in isopropanol before transferring into the load lock chamber and subsequently to the main sputtering chamber. Inside the growth chamber the substrates were slowly heated up to 700 °C for degassing followed by the deposition process. Targets of 99.99% purity of Al and Gd were presputtered for 10-20 mins in a controlled atmosphere composed of 60 % Argon and 40 % $N_2$ before reactively growing AlN and GdN on the substrates. The AlN deposition rate was 1.2 nm/min, at 700 °C**,** with an RF power of 150 W and a working pressure of 1.2 mTorr. The GdN films also were deposited at 700 °C**,** in a working pressure of 2.8 mTorr and a DC power of 50 W. This was followed by another AlN layer for protection. The high-resolution X-ray diffraction (HXRD) pattern shown in Fig. 1(a) displays two peaks, (0002) and (0004), for the AlN, and the (111) and (222) peaks for the GdN film, confirming the preferentially orientated epitaxial growth.

Besides having good crystallinity, it is important that GdN films have a large magnetic moment as confirmed by SQUID measurements, shown in Fig. 1(b). The magnetization saturation of GdN film in the trilayer AlN(10 nm)/GdN($t$)/AlN(10 nm) varies from 5 $\mu_B$/$Gd^{3+}$ to 6 $\mu_B$/$Gd^{3+}$ at 10 K, as the GdN's thickness goes from $t$ = 13 nm to 80 nm, whereas the values are smaller than the theoretical expected value of 7 $\mu_B$/$Gd^{3+}$. The FM critical temperature ($T_C$), was near 30 K, measured with an applied in-plane magnetic field of 50 Oe, as shown in Fig. 1 (c), which agrees with previous reports on thin films of GdN [5, 19-20]. The coercivity ($H_c$) is larger (40 Oe) for the thinner 13 nm-thick GdN film and decreases to ~ 20 Oe for $t$ = 30 and 80 nm.

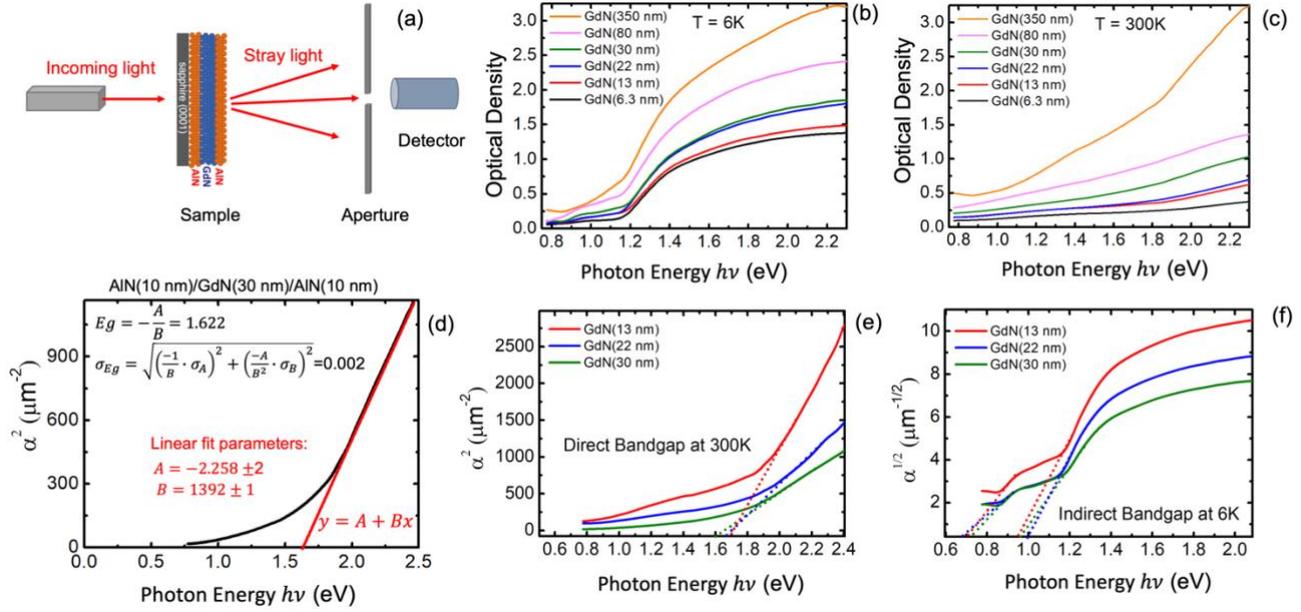

**Figure 2.** (a) Illustration of the optical density acquisition apparatus composed of light source, sample sapphire/AlN/GdN/AlN, aperture and light detector. (b) and (c) show the optical density (OD) spectra for the set of samples at room temperature (300 K) and at 6 K. (d) shows how the direct optical bandgap for a $t = 30$ nm thick film was extracted from the squared-absorption coefficient ($\alpha$) versus photon energy curves using a Tauc plot. The linear red fitted curve ($y = Ax + B$) is used to determine the bandgap. (e) and (f) show Tauc plots for determining the direct and indirect optical bandgaps of the GdN thin films, respectively. The dashed lines indicate the linear extrapolation to $y = 0$ where the intersection corresponds to the direct/indirect bandgap energy.

## 3. Optical Bandgap Properties

Engineering the optical bandgap in GdN thin film by controlling its magnetic state and thickness lead the way for its implementation in spintronic devices. The temperature dependence of semiconductor's bandgap can be different in bulk and thin film morphologies due to factors such as quantum confinement and lattice strain effects. In bulk samples, the bandgap typically increases as the temperature decreases partly due to the renormalization of electron-phonon interactions and partly due to lattice thermal expansions. As the temperature decreases, the phonon population also decreases leading to an increase of the bandgap energy, making it more difficult for electrons to be excited to the conduction band [21]. In most bulk semiconductors, the thermal expansion term is negligible in first approximation compared to the electron–phonon interaction [22]. In contrast, the bandgap's temperature dependence in thin films is more complex. It is known that light absorption by electrons in semiconductors with restricted geometry like thin films, nanowires, and quantum dots are strongly dependent on the size [23-24]. This quantum confinement effect also gives rise to changes in the transition probabilities, which is a consequence of the electron's wave properties. Materials with dimensions on the order of the de Broglie wavelength show relevant quantum-mechanical effects with size-dependent optical properties. Models predict a decrease in phonon frequency and dielectric constant of semiconducting nanostructures leading to an increase of the bandgap energy as the material's size reduces, as confirmed by reported studies [25-27].

On the other hand, the lattice strain plays an important role in the optical properties of thin films leading to changes in the bandgap energy. At low temperatures, the lattice constant of the material shrinks, resulting in an increased interatomic bond strength, higher phonon frequencies and reduced lattice vibrational amplitudes [28]. A compressive strain leads to reductions in bandgap, whereas a tensile strain leads to increases in the bandgap [29]. Besides the lattice parameter changes due to thermal expansion factor, we need to consider the effect of lattice mismatch in multilayered systems. The mismatch promotes extra strain in the interfaces, and thus affects the bandgap. In sapphire/AlN/GdN/AlN, the differences among the thermal expansion coefficients of each material creates complex strain on GdN. GdN lattice expansion coefficient is $8.7 \times 10^{-6}$/K, while for AlN this value is $4.2 \times 10^{-6}$/K, and for sapphire substrate it is $8.1 \times 10^{-6}$/K [30-31]. As the temperature decreases to 6 K, GdN's lattice parameter shrinks more than AlN's, in such a way that the lattice mismatch between GdN and AlN causes a tensile strain on GdN's surfaces, and a compressive strain on AlN's surface, which would lead to an increase in the bandgap of GdN.

| GdN Film Thickness (nm) | A | $\sigma_A$ | B | $\sigma_B$ | $E_g$ | $\sigma_{E_g}$ |
|---|---|---|---|---|---|---|
| 6.3 | -7320 | 30 | 4650 | 20 | 1.576 | 0.009 |
| 13 | -7060 | 20 | 4069 | 7 | 1.736 | 0.005 |
| 22 | -3620 | 20 | 2117 | 7 | 1.709 | 0.009 |
| 30 | -2258 | 2 | 1392 | 1 | 1.622 | 0.002 |
| 80 | -388.8 | 0.8 | 261.0 | 0.4 | 1.490 | 0.004 |
| 350 | -229.2 | 0.9 | 137.6 | 0.4 | 1.665 | 0.008 |

**Table I.** Intercept (A) and linear fit slope (B) and their respective uncertainties ($\sigma_A$ and $\sigma_B$) from Tauc plots for the AlN/GdN(*t*)/AlN structures as a function of GdN thickness (*t*). The direct bandgap (*Eg*) and its uncertainty ($\sigma_{E_g}$) for GdN films at 300 K in eV units.

To better understand the role of thickness and temperature in a thin GdN layer, we measured the optical density (**OD**) spectra as a function of the photon energy ($E = h\nu$), at temperatures of 300 K and 6 K, for AlN(10 nm)/GdN(*t*)/AlN(10 nm) trilayers on (0001)-sapphire with *t* = 6.3, 13, 22, 30, 80 and 350 nm (see Figs. 2 (b) and (c)). The absorption coefficient ($\alpha$) was determined using the Beer-Lambert law that relates the amount of light absorbed by a material to its concentration, path length, and molar absorption coefficient. For a material with a uniform thickness, the relationship can be simplified as:

$$\alpha = \frac{OD}{t}, \qquad (1)$$

where *t* is the film's thickness. Tauc plots were used to estimate the direct bandgap ($E_g$) energies of the GdN thin films, which consists in plotting $\alpha^2$ versus $E$ and then, extrapolating the linear portion of the curve to $\alpha^2 = 0$, as shown in Fig. 2(d), where $E_g = 1.6$ eV. The energy at which the extrapolated line intersects the energy axis corresponds to the direct bandgap energy $E_g$ [32-34]. Note that this extrapolated value is considerably larger than the energy where the absorption approaches zero, as has been used in the past to identify the bandgap energy. The extrapolation is more accurate as it reduces the effects of band broadening and the possible appearance of impurity states lying below the band edges. To determine the indirect bandgap, the procedure is similar, but the analysis relies on extrapolating $\alpha^{1/2}$ versus $E$. It is important to note that the range of the photon energy in this investigation is well below that for the AlN bandgap of ~ 6.1 eV and sapphire bandgap of ~ 9 eV [35-37]. Thus, the observed absorption spectra are clearly attributable to the GdN films. Figures 2 (e) and (f) show Tauc plots at 300 K and 6 K, respectively. To reduce errors in the bandgap determination due to drawing lines manually, we identified the linear portions of each curve ($\alpha^2$ versus $E$) and fit the experimental data to $y = A + Bx$. The bandgap was determined when $y = \alpha^2 = 0$, so $E_g = -A/B$. The uncertainty associated with this bandgap was determined using:

$$\sigma_{E_g} = \sqrt{\left(\frac{-1}{B}\cdot\sigma_A\right)^2 + \left(\frac{-A}{B^2}\cdot\sigma_B\right)^2}, \qquad (2)$$

where $\sigma_A$ and $\sigma_B$ are the relevant uncertainties. Table 1 shows fitted values of the intercept (A) and slope (B) and their respective uncertainties that were used for determining the direct bandgap at 300 K. The red data points in Fig. 3 (d) show $E_g(t)$, where the solid curve is a guide for eye and the error bars are smaller than the data points. The direct bandgap of the GdN films in the paramagnetic phase is around $E_{300K} = 1.6$ eV and spans a region of GdN thickness (13 nm < *t* < 80 nm) where the gap increases as the thickness decreases, being attributed to quantum confinement. When samples are cooled down to 6 K they become FM, where the absorption spectra present more than one narrow linear deviation, indicating a spin split bandgap. Thus, in the magnetic state, the average minority (spin-down) bandgap energy is $E_\downarrow \approx 1.2$ eV, and the average majority (spin-up) bandgap energy is $E_\uparrow \approx 0.8$ eV, as shown by the blue data in Fig. 3 (d), where the black dotted lines correspond to their average energy values. As the temperature decreases from 300 to 6 K, the bandgap variations are $E_{300K} - E_\downarrow \approx 0.4$ eV for the minority state and $E_{300K} - E_\uparrow \approx 0.8$ eV for the majority state. This observed bandgap reduction as the temperature decreases to 6K indicates the shrinking of the lattice parameter in the volume of GdN overcomes the tensile strain effect at the interface of AlN/GdN that would lead to an increase in the bandgap.

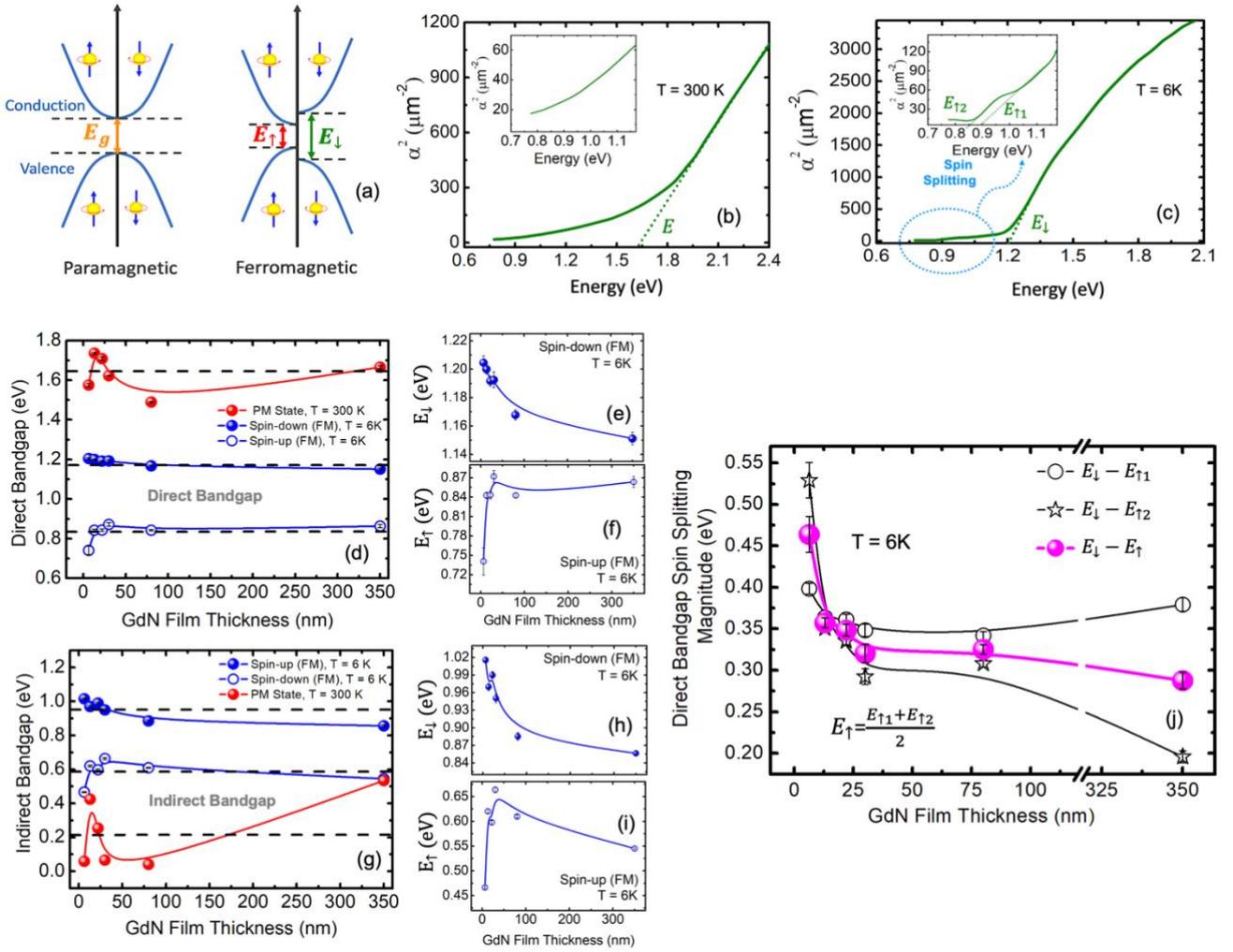

**Figure 3.** (a) Illustration of the change in the band structure of a magnetic semiconductor depending on the electron's spin orientation in the paramagnetic (left) and ferromagnetic (right) phases. In the PM phase, the bandgap doesn't depend on the spin's orientation. (b) and (c) compare the Tauc plots for obtaining the direct optical bandgap of sapphire/AlN(10 nm)/GdN(32 nm)/AlN(10 nm) in the PM ($T = 300$ K) and FM ($T = 6$ K) phases. In the FM phase, shown in the inset of (c), we observe more than one linear portion that is related to the spin split bandgap. (d) and (g) show the direct and indirect optical bandgap dependence with the GdN film thickness at 300 K and 6 K, where the black dotted lines correspond to their average energy values, and (e), (f), (h) and (i) are the zoomed direct and indirect bandgaps for the minority and majority spin states. The spin splitting of the bandgap is also present in the indirect transitions. (j) Dependence of the magnitude of the spin-splitting as a function of the GdN thickness, showing that the splitting becomes larger as the film thickness decreases.

Finally, Figure 3 (a) is an illustration of the band structure of a semiconducting magnetic material in its paramagnetic and ferromagnetic phases. The bandgap in PM phase is the same for electrons with spin-up or spin-down, whereas in the FM phase it depends on the spin orientation, and it splits into $E_↓$ and $E_↑$. Figures 3 (b) and (c) demonstrate Tauc's plots for the direct bandgap of a 30 nm thick GdN film in both paramagnetic (300 K) and ferromagnetic (6 K) phases, respectively. In the FM state, we observe a linear portion that extrapolates to $E_↓ = 1.192 ± 0.006$ eV, and by zooming around 0.9 eV, as shown on the inset, we observe two linear portions that extrapolate to the energies of $E_{↑2} = 0.844 ± 0.005$ eV and $E_{↑1} = 0.900 ± 0.008$, with an average value of $E_↑ = 0.87 ± 0.01$. This bandgap splitting appears for all GdN thicknesses investigated here, and a complete analysis is shown in Fig. 3 (d). The filled blue points correspond to the minority bandgap ($E_↓$) and the unfilled blue points correspond to the majority bandgap ($E_↑$). The details are shown in the zoomed Figures 3 (e) and (f), respectively. These results show a giant optical spin splitting of the direct bandgap at 6 K of $E_↑ − E_↓ ≈ 0.4$ eV. The dependence of the direct bandgap spin splitting magnitude on the GdN film thickness is shown in Fig. 3 (j). The black star and black circle points show the splitting with respect to $E_{\uparrow 1}$ and $E_{\uparrow 2}$, and the solid lines are guides for the eyes. As the thickness decreases, the spin-splitting magnitude increases most likely due to two combined factors: (i) The strengthening of the quantum confinement as GdN film becomes thinner; and (ii) The reduction of the GdN's volume making the surface tensile strain on GdN more relevant, thus leading to a bandgap increase.

Figure 3 (g) shows the same analysis for the indirect optical bandgaps in the PM (red points) and FM states (blue filled and unfilled points). At room temperature (300 K) the indirect transitions occur at very small energies, whereas in the FM phase, we still observe two absorption lines with separation of $\sim 0.4$ eV that increases as film's thickness decreases reaching values of $\sim 0.6$ eV for the 6.3 nm thick layer. The zoomed details are shown in Figs. 3 (h) and (f). This systematic investigation shows that it is possible to control the bandgap of GdN by controlling growth conditions and magnetic state and facilitates the implementation of this promising material in complex devices to act as spin filters and spin injectors.

## 4. Conclusions

We investigated a set of epitaxial GdN thin films aiming to control its bandgap and move towards their application as a spin-polarized source in spintronics devices such as spin transistors, TMR magnetic tunnel junctions and spin valves. Our results show that it is possible to manipulate the bandgap by selecting an appropriate film thickness and magnetic phase. By varying the thickness from 6 nm to 350 nm we observed a room temperature direct bandgap average value of 1.6 eV with a variation of $0.24$ eV as the thickness decreases. Upon cooling the films down to 6 K, in their ferromagnetic phase the direct bandgap splits into two gaps of 1.2 eV to 0.8 eV, corresponding to the minority and majority spin bands. This giant optical spin-splitting of $0.4$ eV in the direct optical bandgap for the ferromagnetic state confirms this material is suitable for spin-filtering applications. The same behavior appears in the indirect bandgap where we observed a splitting magnitude of $0.3$ eV. An important takeaway is that the spin splitting for both bandgaps becomes larger for thinner films, which is attributed to quantum confinement and enhancement of the surface tensile strain due to lattice mismatch between AlN/GdN layers. The observed spin splitting magnitude ($E_{\uparrow} - E_{\downarrow}$) increased 60% as the GdN thickness varied from 350 nm to 6 nm. This remarkable result sheds light on the possibility of manipulation and control of the optical and magnetic properties of the semiconducting GdN thin films with potential for high spin polarization through the spin splitting of the bandgap. Further investigations should focus on determining the bandgap dependence on a larger range of temperature covering the details on the PM-FM transition. Besides that, it would be interesting to perform X-ray measurements as a function of temperature to understand the role of thermal expansion and lattice mismatch on the bandgap of GdN thin films.

## 5. Acknowledgements

This research is supported by ARO (W911NF1920041), NSF (DMR 1700137, DMR 2218550), ONR (N00014-16-1-2657, N00014-20-1-2306), the Center for Integrated Quantum Materials (NSF-DMR 1231319) and Brazilian agencies CAPES (Gilvania Vilela/POS-DOC-88881.120327/2016-01), FACEPE (APQ-0565-1.05/14), CNPq and UPE(PFA/PROGRAD/UPE 04/2017). Work at Northeastern University was partially supported by the National Science Foundation (USA) grant DMR-1905662 and the Air Force Office of Scientific Research (USA) award FA9550-20-1-0247. Work at São Paulo University was supported by FAPESP (grant 2022/15791-2).

## 6. References

[1] F. Natali, B. J. Ruck, N. O. V. Plank, H. J. Trodahl, S. Granville, C. Mayer, and W. R. L. Lambrecht, Rare-earth mononitrides. Prog. Mater. Sci. **58**, 1316 (2013); https://doi.org/10.1016/j.pmatsci.2013.06.002

[2] C. M. Aerts, P. Strange, M. Horne, W. M. Temmerman, Z. Szotek, and A. Svane, Half-metallic to insulating behavior of rare-earth nitrides. Phys. Rev. B **69**, 045115 (2004); https://doi.org/10.1103/PhysRevB.69.045115

[3] Ilia N. Sivkov, Oleg O. Brovko, and Valeri S. Stepanyuk, Spin-polarized transport properties of GdN nanocontacts. Phys. Rev. B **89**, 195419 (2014); https://doi.org/10.1103/PhysRevB.89.195419

[4] Pal, A., Senapati, K., Barber, Z.H. and Blamire, M.G. (2013), Electric-Field-Dependent Spin Polarization in GdN Spin Filter Tunnel Junctions. Adv. Mater. **25**: 5581-5585 (2013); https://doi.org/10.1002/adma.201300636

[5] H. Yoshitomi, S. Kitayama, T. Kita, O. Wada, M. Fujisawa, H. Ohta, and T. Sakurai, Optical and magnetic properties in epitaxial GdN thin films. Phys. Rev. B **83**, 155202 (2011); https://doi.org/10.1103/PhysRevB.83.155202

[6] Ilia N. Sivkov, Oleg O. Brovko, and Valeri S. Stepanyuk, Spin-polarized transport properties of GdN nanocontacts.

Phys. Rev. B **89**, 195419 (2014); https://doi.org/10.1103/PhysRevB.89.195419

[7] Chun-gang Duan, R. F. Sabiryanov, Jianjun Liu, and W. N. Mei, Strain Induced Half-Metal to Semiconductor Transition in GdN. Phys. Rev. Lett. **94**, 237201 (2005); https://doi.org/10.1103/PhysRevLett.94.237201

[8] S. Granville, B. J. Ruck, A. Koo et al., Semiconducting ground state of GdN thin films. Phys. Rev. B **73**, 235335 (2006); https://doi.org/10.1103/PhysRevB.73.235335

[9] P. Wachter, Physical properties of stoichiometric GdN single crystals. Results in Physics **2**, 90–96 (2012); http://dx.doi.org/10.1016/j.rinp.2012.07.003

[10] Tiffany S. Santos and Jagadeesh S. Moodera, Spin Filter Tunneling. Spintronics Handbook, 2nd Edition: Spin Transport and Magnetism, CRC Press (2019). ISBN 9780429423079

[11] R. Vidyasagar, S. Kitayama, H. Yoshitomi, T. Kita1, T. Sakurai, and H. Ohta, Study on spin-splitting phenomena in the band structure of GdN. Appl. Phys. Lett. **100**, 232410 (2012); https://doi.org/10.1063/1.4727903

[12] R. Vidyasagar, T. Kita, T. Sakurai, and H. Ohta, Giant optical splitting in the spin-states assisting a sharp magnetic switching in GdN thin films. Appl. Phys. Lett. **102**, 222408 (2013); https://doi.org/10.1063/1.4809758

[13] Muhammad Azeem, Spin-Split Joint Density of States in GdN. Chin. Phys. Lett. **33** 92), 02750 (2016); https://iopscience.iop.org/article/10.1088/0256-307X/33/2/027501

[14] D. Massarotti a, R. Caruso, A. Pal, G. Rotoli, L. Longobardi, G.P. Pepe, M.G. Blamire, F. Tafuri, Low temperature properties of spin filter NbN/GdN/NbN Josephson junctions. Physica C: Supercond. Applic. **533**, pages 53-58 (2017); https://doi.org/10.1016/j.physc.2016.07.018

[15] Ahmad, H.G., Minutillo, M., Capecelatro, R. et al. Coexistence and tuning of spin-singlet and triplet transport in spin-filter Josephson junctions. Commun. Phys 5, 2 (2022). https://doi.org/10.1038/s42005-021-00783-1

[16] R. J. Gambino, T. R. McGuire, H. A. Alperin, and S. J. Pickart, Magnetic Properties and Structure of GdN and $GdN_{1-x}O_x$. J. Appl. Phys. **41**, 933 (1970); https://doi.org/10.1063/1.1659030

[17] Stefan Cwik, Sebastian M. J. Beer, Stefanie Hoffmann et al. Integrating AlN with GdN Thin Films in an in Situ CVD Process: Influence on the Oxidation and Crystallinity of GdN. ACS Appl. Mater. Interfaces, **9**, 27036−27044 (2017); https://doi.org/10.1021/acsami.7b08221

[18] H. J. Trodahl, A. R. H. Preston, J. Zhong, and B. J. Ruck, Ferromagnetic redshift of the optical gap in GdN. Phys. Rev. B **76**, 085211 (2007); https://doi.org/10.1103/PhysRevB.76.085211

[19] R. Vidyasagar, T. Kita, T. Sakurai, and H. Ohta, Electronic transitions in GdN band structure. J. Appl. Phys. **115**, 203717 (2014); https://doi.org/10.1063/1.4880398

[20] F. Leuenberger, A. Parge, W. Felsch, K. Fauth, and M. Hessler, GdN thin films: Bulk and local electronic and magnetic properties. Phys. Rev. B **72**, 014425 (2005); https://doi.org/10.1103/PhysRevB.72.014427

[21] Y. P. Varshni, Temperature dependence of the energy gap in semiconductors, Physica **34**, 149 (1967); https://doi.org/10.1016/0031-8914(67)90062-6

[22] P.B. Allen and M. Cardona, Temperature dependence of the direct gap of Si and Ge, Phys. Rev. B **27**, 4760 (1983); https://doi.org/10.1103/PhysRevB.27.4760

[23] Gaponenko, S. and Demir, H., Quantum confinement effects in semiconductors, Applied Nanophotonics (pp. 52-91). Cambridge: Cambridge University Press. https://doi.org/10.1017/9781316535868.004

[24] P. Dey, J. Paul, J. Bylsma, D. Karaiskaj, J.M. Luther, M.C. Beard, A.H. Romero, Origin of the temperature dependence of the band gap of PbS and PbSe quantum dots, Sol. State Comm. **165**, 49 (2013); http://dx.doi.org/10.1016/j.ssc.2013.04.022

[25] M. Li and J.C. Li, Size effects on the band-gap of semiconductor compounds, Materials Letters 60, 2526–2529, (2006); https://doi.org/10.1016/j.matlet.2006.01.032

[26] M Singha, BM Taele, M Goyal, Modeling of size and shape dependent band gap, dielectric constant and phonon frequency of semiconductor nanosolids, Chinese J. Phys. **70**, 26–36 (2021); https://doi.org/10.1016/j.cjph.2021.01.001

[27] A. Barnasas et al., Quantum confinement effects of thin ZnO films by experiment and theory, Physica E **120**, 114072 (2020); https://doi.org/10.1016/j.physe.2020.114072

[28] B Bertheville, H Bill and H Hagemann, Experimental Raman scattering investigation of phonon anharmonicity effects in $Li_2S$, J. Phys.: Condens. Matter **10**, 2155 (1998); https://dx.doi.org/10.1088/0953-8984/10/9/018

[29] D. A. Shohonov et al., Effects of lattice parameter manipulations on electronic and optical properties of BaSi2,

Thin Sol. Films **686**, 137436 (2019); https://doi.org/10.1016/j.tsf.2019.137436

[30] Gyeonghun Kim and Sangjoon Ahn, Thermal conductivity of gadolinium added uranium mononitride fuel pellets sintered by spark plasma sintering, J. Nuc. Mater. **546**, 15278 (2021); https://doi.org/10.1016/j.jnucmat.2021.152785

[31] W. M. Yim and R. J. Paff, Thermal expansion of AlN, sapphire, and silicon, J. Appl. Phys. **45**, 1456 (1974); https://doi.org/10.1063/1.1663432

[32] J. Tauc, Optical Properties and Electronic Structure of Amorphous Ge and Si. Mat. Res. Bull. Vol. 3, pp. 37-46(1968). Pergamon Press, Inc. Printed in the United States.

[33] P.R. Jubu, F.K. Yam, V.M. Igba, K.P. Beh, Tauc-plot scale and extrapolation effect on bandgap estimation from UV–vis–NIR data – A case study of β-$Ga_2O_3$. J. Sol.State Chem. **290**, 121576 (2020); https://doi.org/10.1016/j.jssc.2020.121576

[34] Brian D. Viezbicke, Shane Patel, Benjamin E. Davis and Dunbar Birnie, Evaluation of the Tauc Method for Optical Absorption Edge Determination: ZnO Thin Films as a Model System. Physica Status Solidi, B, **252**(8), 1700-1710 (2015); https://doi.org/10.7282/T3W097T7

[35] Xiao Tang, Fazla Hossain, Kobchat Wongchotigul, and Michael G. Spencer, Near band-edge transition in aluminum nitride thin films grown by metal organic chemical vapor deposition. Appl. Phys. Lett. **72**, 1501 (1998); https://doi.org/10.1063/1.121039

[36] J. A. Pérez Taborda, H. Riascos Landázuri and L. P. V. Londoño, Correlation Between Optical, Morphological, and Compositional Properties of Aluminum Nitride Thin Films by Pulsed Laser Deposition. IEEE Sensors Journal **16**, (2), 359-364 (2016), https://doi.org/10.1109/JSEN.2015.2466467

[37] Alang Kasim Harman, Susumu Ninomiya, and Sadao Adachi, Optical constants of sapphire (α-Al2O3) single crystals. J.Appl. Phys. **76**, 8032 (1994); https://doi.org/10.1063/1.357922